\documentclass{sig-alternate}

\begin{document}\sloppy






\title{TopoMan: Global Network Visibility in the Presence of Middleboxes (A Graybox Approach)} 

\numberofauthors{1} 
\author{
Vasudevan Nagendra, Shubhada Patil, Michalis Polychronakis, Samir R. Das \\
        \affaddr{Stony Brook University, NY}\\
      	\email
       {\{ vnagendra, sppatil, smikepo, samir    \}@cs.stonybrook.edu}
}

\maketitle
\begin{abstract}
Software Defined Networks (SDN) provide vital benefits to network administrators by offering \textit{global visibility} and \textit{network-wide control} over the switching infrastructure of the network. It is rather much difficult to obtain the same benefits in the presence of middleboxes (MBs), due to (i) lack of a proper topology discovery mechanism in environments with a mix of forwarding devices and middleboxes. (ii) lack of generic APIs to abstract and gain control on these rigid and heterogeneous third-party middleboxes (iii) lack of a generic network infrastructure framework to monitor and verify any specific device or path connectivity status in the network. These limitations make automation of network operations such as, network-wide monitoring, policy enforcement and rule-placement much difficult to handle. Hence, there is a greater urge even from middlebox vendors, to better handle the \textit{control} and \textit{visibility} aspects \cite{ietf-draft-gu-statemigration-framework-03,draft-ietf-midcom-protocol-rfc4097,draft-ietf-midcom-rfc5189} of the network in presence of middleboxes.

In this paper, we propose a \textit{Unified network infrastructure framework} for gaining \textit{global network visibility}, by discovering the network topology in the presence of middleboxes, along with a framework to support the end-to-end path connectivity verification, \textit{independent of SDN}. We have also addressed security aspects and provided necessary APIs to support our framework.


\end{abstract}

%
%




\section{Introduction}
\noindent \newline
\noindent Today's enterprise networks are proliferating with vast number of specialized devices or middleboxes \cite{Sherry:EECS-2012-24,Sekar:2011:MME:2070562.2070583}, indeed the middlebox market is considered as one of the fastest growing market segments with Compound Annual Growth Rate(CAGR) \cite{CAGR_NetSec2019} of 11\%. Even with such a rapid growth rate and maturity in middlebox market, we still lack a \textit{standard interface} to control these devices or a \textit{framework to unify} them. Hence, gaining \textit{global network visibility} in the presence of middleboxes is considered as one of biggest challenges in the network. 

Consider for example, enforcing \textit{Network Function Chain} rules on to network flows requires the administrator to calculate the candidate paths and identify right devices (switches and middleboxes) to place the rules. In dynamic networks of huge size, such network management solutions can not be automated with out having access to complete network topology. Also, the different middlebox deployment mechanisms adapted in today's enterprise networks, such as inline (i.e., output of one middlebox is directly connected to input of other middle box), offline (i.e., middleboxes doesn't fall in the active path of the traffic, the traffic is steered to them using forwarding devices) cluster and tree topologies makes automatic topology discovery a complicated task.

The existing discovery protocols such as LLDP/BDDP \cite{7021050} can not be used directly with middleboxes, as they are limited only to ethernet segment. Also, these protocols have limitations discovering devices with TAP, virtual and VLAN interfaces, and are not compatible with VPN and other overlay technologies. Other protocols such as SNMP (Simple Network Management Protocol) and MIDCOM \cite{MIDCOM} are configuration based device management protocols, lacking capability to automatically discover the topology of middleboxes. Tracebox \cite{Detal:2013:RMI:2504730.2504757}, TCPM \cite{TCPMOption}, MIDAS \cite{midas} and RSVP \cite{RSVP} could only be used in determining the middleboxes along a specific path of the network. All these mechanisms requires a separate traffic generating entity to initiate special requests for discovering the middleboxes along the path. But, obtaining such vantage points to generate such traffic to extract the path details might not always be feasible in network deployments. Recent work HybridSDN \cite{hongincremental} proposed a topology discovery mechanism in hybrid enterprise networks using dynamic routing protocols, by completely discarding the middleboxes. Most importantly, the discovery protocols discussed above lacks security and are prone to visibility poisoning attacks \cite{hong2015poisoning} and data leaks. Hence, there is strong need for an \textit{effective} and \textit{secured} topology discovery mechanism in the presence of middleboxes. Importantly, we utilize the capabilities of middleboxes itself to generate the probes without using other special devices as suggested in other approaches \cite{Detal:2013:RMI:2504730.2504757,TCPMOption,midas,RSVP}. Our main contributions in this position paper includes obtaining better visibility over the enterprise networks by building,

\begin{itemize}

\item 
a probe-based \textit{secured} topology discovery mechanism for middleboxes, independent of SDN. 

\item
probe-based framework for end-to-end path connectivity verification.

\item
generic middlebox APIs to support our topology discovery framework.

\end{itemize}

\section{SYSTEM ARCHITECTURE}
\begin{figure}[]
\includegraphics{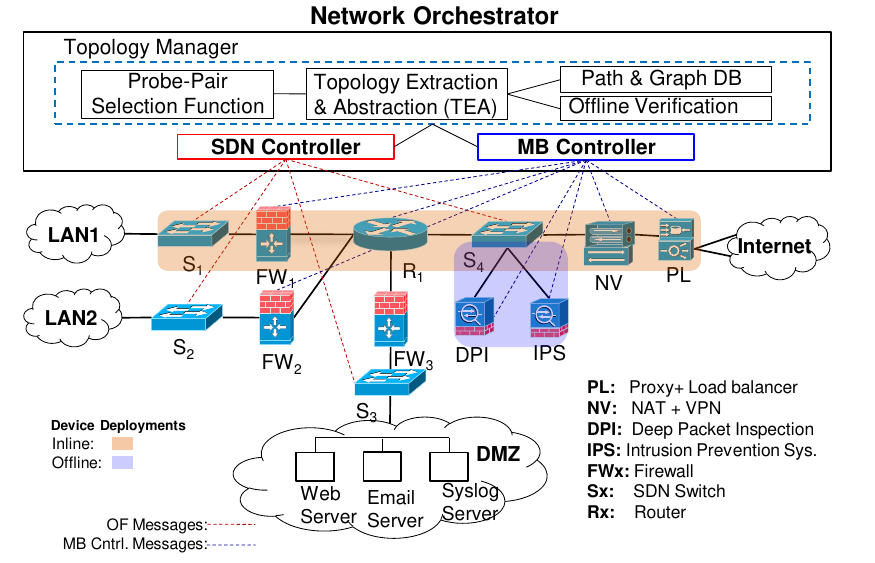}
\caption{\small Unified enterprise network architecture.}
\label{figure: topoman2}
\end{figure}

\noindent \newline
\noindent The Unified network infrastructure framework shown in figure \ref{figure: topoman2} uses a simple text based application protocol (shown in figure \ref{figure: Probe protocol details}) for topology discovery mechanism. In this approach, the probe message is generated between set of middleboxes and it is allowed to traverse the network, and these traversal details are used to build the topology of middleboxes. This process is repeated until the complete middlebox topology is discovered. The intricate details of this probe-based topology discovery mechanism is discussed in section \ref{topology discovery}. The main functional components of this high level system architecture is detailed below,

\subsection{Topology Manager} 
This module has three functional sub-components i.e., \textit{Topology Extraction and Abstraction module (TEA)}, \textit{Probe-pair Selection Function (PSF)} and \textit{Offline Verification} module to help build the topology and verify it. The topology Manager consolidates the topology details obtained from both SDN and MB Controllers to build the complete network topology.

\noindent \textbf{Topology Extraction \& Abstraction.} This module basically performs two different functions, \textit{topology discovery} and \textit{end-to-end path connectivity verification}. It devise a means to select and optimize the right set of probe-pairs\footnote{Pair of middleboxes, a source \& destination middlebox between which the probe traverses to discover the topology.} required to discover the network topology. The topology manager triggers source middlebox (MB Agent) of the probe-pair to generate a probe message and forwards it further to destination middlebox of the probe-pair. This module collects the reports sent by the middleboxes along the path of probe traversal to incrementally build the topology. TEA module also triggers a special type of probe message for checking the path connectivity.

\noindent \textbf{Offline Verification.} This module helps \textit{TEA} to verify the topology built through probe-based discovery mechanism and report discrepancies if any.

\noindent \textbf{Probe-pair Selection Function.} This helper function uses heuristics for selecting the right probe-pair combination for topology discovery, as explained in section \ref{probe pair selection}.

\subsection{MB Agent} 
The MB Agent has following functional sub-components, \textit{Transparent Proxy Based Agent (TPA)}, \textit{Probe Handler (PH})  and \textit{Probe Tracking \& Control (PTC}) module. The high level middlebox architecture of MB Agent is shown in figure \ref{figure: middlebox architecture}.

\noindent \textbf{Transparent Proxy based Agent.} TPA is a transparent proxy module which digests, generates and forwards the probes traversing through it. It also acts as an interface between the middlebox and MB Controller.

\noindent \textbf{Probe Handler.} This module handles both the discovery and path connectivity verification probes and forwards them depending on type of the probe message.

\noindent \textbf{Probe Tracking and Control.} PTC steers the probe message on any specific output interface as guided by the MB Agent, which also helps in monitoring the outgoing probe message.

\subsection{Topology Discovery Probes}

We use two types of probe messages, one for topology discovery (\textit{Discovery probe}) and other for path connectivity verification (\textit{path checker probe}). The \textit{path-id} field in the probe header and the port number differentiates these two types of probe messages. 

\subsection{Middlebox APIs and Messages} \label{mb apis}

MB APIs are used for interfacing the communication between MB Agents and MB controller (\textit{Topology Manager}). 
The important APIs used in our prototype are, \textit{{\small DEVICE-CAPABILITIES}} used to exchange the details of each of the middlebox with MB Controller, \textit{{\small PROBE-INIT}} to trigger the MB Agent to initiate a probe message to destination middlebox with specific security header flags, \textit{{\small PROBE-UPDATE}} to update the received probe message details to the MB Controller, \textit{{\small UPDATE-OUTINTERFACE}} to update the corrected output interface of probe to the MB Controller, \textit{{\small RESOLVE-PROBEID}} to resolve the Random Number to Probe-pair IP addresses when header security is enabled, \textit{{\small HEARTBEAT}} to update the device status to the MB Controller.

\section{Topology Discovery} \label{topology discovery} 
\begin{figure}
\includegraphics{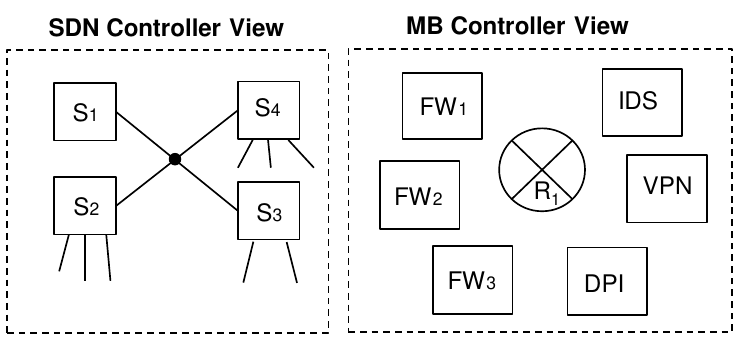}
\caption{\small Initial phase network view.}
\label{figure: global view initial}
\vskip -5pt
\end{figure}

\noindent \newline
\noindent As part of topology discovery mechanism, middleboxes i.e., MB Agents initially exchanges their \textit{device-capabilities} with MB Controller (Topology Manager). The consolidated details at the \textit{Topology Manager} after this Initial capabilities exchange phase is shown in figure \ref{figure: global view initial}, as list of discrete middlebox nodes and SDN topology Islands (if available). For exposing the device capabilities we are using \textit{Graybox model}, rather than using \textit{Whitebox model} requiring the middlebox vendors to expose \textit{Full state machine (FSM)} details, which vendors will not be comfortable exposing.

\subsection{Probe-based Topology Discovery}\label{probe-based topo discovery} 
The topology manager selects a probe-pair from list of available middleboxes after the initial {\small DEVICE-CAPABILITIES} exchange. It then sends {\small PROBE-INIT} message to trigger the source MB Agent to initiate a probe to its destination middlebox, as illustrated in figure \ref{figure: probe traversal}. The details of the probe-pair selection is mentioned in the section \ref{probe pair selection}. As most of the middleboxes are multi-homed devices, the middlebox agent chooses suitable interface of the selected probe-pair to generate the traffic between those interfaces of the Probe-pair.

\begin{figure}
\centering
\includegraphics{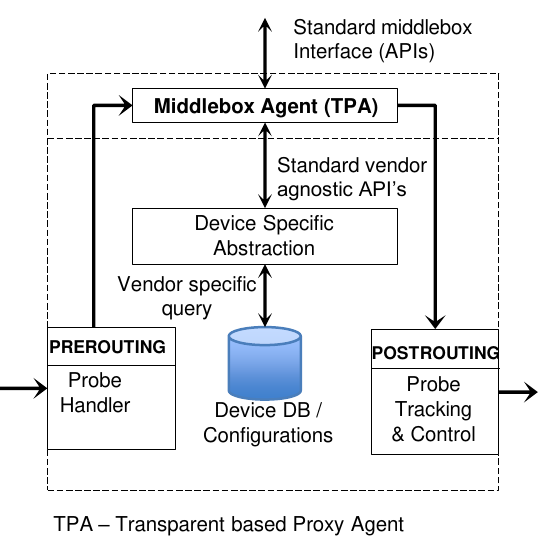}
   \caption{\small High level middlebox architecture.} 
   \label{figure: middlebox architecture}
\vskip -5pt

\end{figure}

The MB Agent of the source middlebox generates probes to all the interface IP addresses of the destination probe-pair, only if there are alternate routes available to the destination middlebox interfaces from the source middlebox. 
When the probe message encounters a SDN switch in its path, the probe details are exchanged with the SDN Controller (floodlight) and then the SDN Controller forwards it out of the switch through a Packet-out message. Similarly, when the probe message encounters a middlebox in its path, the {\small PREROUTING}\footnote{Handling Ingress packets before routing decision.} \textit{Probe-Handler} module in the middlebox as shown in figure \ref{figure: middlebox architecture} takes the probe message up the layer to MB Agent by using redirect rules. The MB Agent then digests this probe message and further forwards it out of the middlebox. The {\small POSTROUTING}\footnote{Handling Egress packets after routing decision.} \textit{Probe Tracking and Control} module tracks the probe messages going out of the device and update the details to MB Agent, which indeed exchanges these details to Topology Manger through MB controller. 

\noindent \textbf{Populating the Probe Payload.} Before sending the probe out of middlebox, it is populated with the current device details along with its input and output interfaces the probe has taken. Calculating the output interface details of probe, even before it is sent out is a difficult task, as certain devices like load-balancers might choose different output interface dynamically. 


We handle this situation using two different approaches, (i) The output interface is calculated using the route details of the device and updated to the probe payload. Later, the calculated output interface value is compared with the actual output interface taken by the probe and the corrected value is updated to the Topology Manager using {\small UPDATE-OUTINTERFACE API}. 
(ii) In the other approach, rather than using default route policies, the MB Agent statically calculates the output interface for destination IP of the probe and steers it out on that specific interface using the \textit{PTC} module.

\noindent \textbf{Probe Discovery Termination.} 
The topology manager maintains \textit{discovered and undiscovered interfaces list} of all middleboxes in the network and the probe-pairs are always selected from the \textit{undiscovered} list. Once all the interfaces of a middlebox are discovered (i.e., links to their adjacent nodes) through our probe-based discovery mechanism, it is moved from 'undiscovered list' to 'discovered list'. The topology discovery process terminates when the middleboxes in the undiscovered list are either completely exhausted or left out with only edge-middlebox interfaces for which the link connectivity couldn't be directly discovered. We could easily identify these interfaces in advance before the discovery process itself, using approaches mentioned in section \ref{probe pair selection}. Also, importantly we will not iterate the discovery process over same probe-pair interface more than twice to avoid discovery mechanism from running in to loops.

\begin{figure}[t]
\centering
\includegraphics{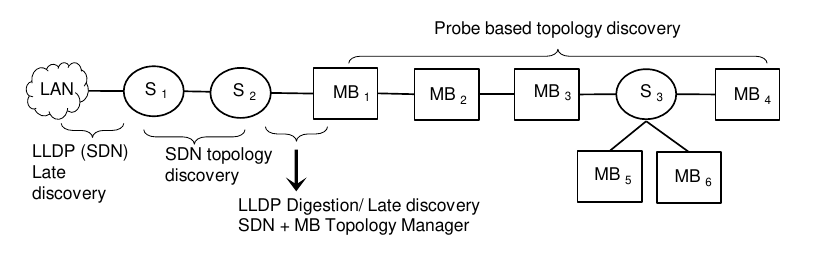}
\caption{\small Overall topology discovery mechanism.}
\label{figure: topodisc}
\end{figure}

\noindent \textbf{Late Discovery.} In networks with mix of SDN switches and middleboxes, certain interface links of middleboxes that are directly connected to edge SDN switches are left out from discovery, which are discovered lately when actual traffic starts flowing through that interface as illustrated in figure \ref{figure: topodisc}.

\subsection{Probe-pair selection \& heuristics} \label{probe pair selection}
In our approach, we use following heuristics to automatically identify the edge-devices in the network as probe-pairs. In general, the edge devices are those which lie in the east-west boundaries of the network, when chosen as probe-pairs they are expected to discover more number of devices along its path effectively reducing the number of probes or up-calls needed to build the topology. 

\noindent \textbf{Heuristic 1: Interface based probe-pair selection.} Certain devices can automatically be identified as edge device depending on their interface sub-nets, i.e., a device is considered as edge if none of it's interface IP addresses falls under the same IP sub-net of any other middleboxes in the network. 

\noindent \textbf{Heuristic 2: Policy based probe pair selection.} In this approach, we used network policies (policy configuration file), to extract edge devices in the network. This is achieved by grouping the network policies on the basis of their source and destination networks and comparing with the interface IP addresses of the middleboxes, we were able to identify the edge devices by matching their networks. Though these heuristics are not any standard approaches, but, still it reduced the probes inside network by around 10 to 35\%.  

\subsection{End-to-end path connectivity Verification}

In general, finding end-to-end path connectivity inside a complex network for a specific traffic or service is not a straight forward task. It requires the administrator to have vantage points in the network for generating test traffic, and also requires a means to manually verify all the devices along the path. 

\begin{figure}
\centering
\includegraphics{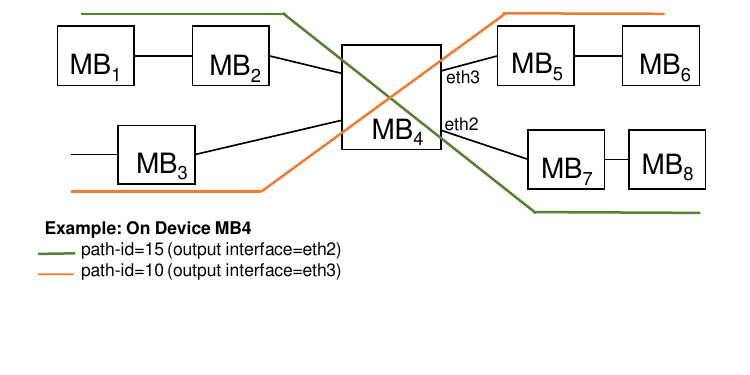}
\caption{\small Probe-based path connectivity checker.}
\label{figure: probepath}
\end{figure}

\noindent \textbf{Approach.} With our approach, the \textit{Topology Manager} pre-configures all the devices in the path (to be verified) with port based(transport) routing rules depending on the path-id. After this initial path setup is done, the \textit{Topology Manager} triggers the source middlebox (MB Agent) to generate a path checker probe message with a specific path-id. The middleboxes forwards this probe message on the basis of the path-id, which is illustrated in the figure \ref{figure: probepath}. The \textit{Probe-Handler} and \textit{PTC }modules present in each of the middlebox uses the port based routing to steer this probe message through the specified path pertaining to the path-id.

\subsection{Reduce Up-call count (Probe-append)} 
\begin{figure}[t]
\centering
\includegraphics{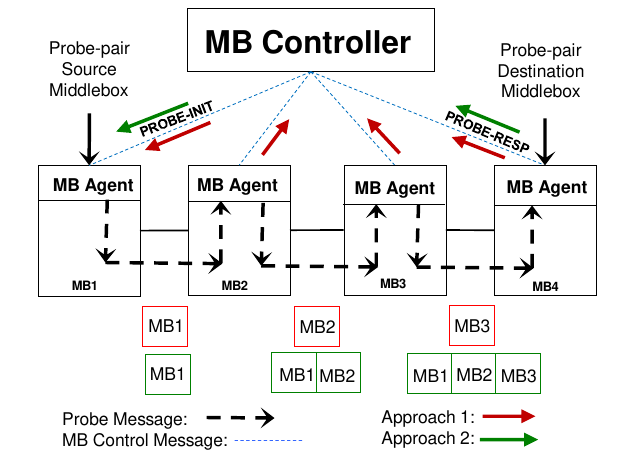}
\caption{\small Flow diagram of probe message traversal.}
\label{figure: probe traversal}
\end{figure}

In contrast to existing approach (Approach 1) as illustrated in figure \ref{figure: probe traversal}, rather than making an up-call\footnote{Message to MB Controller updating the probe details.} at each node of probe traversal, the payload of the probe message is appended with the current node details. This process of payload appending is carried out all the way till probe reaches the destination probe-pair middlebox, where an up-call is finally made to the MB controller as shown in figure \ref{figure: probe traversal} (Approach 2). The only exception to this probe handling approach is, when the probe encounters a SDN switch in the middle of its traversal, it simply forwards the probe further.

\subsection{Static topology verification (Simulator)} \label{simulator}

This tool is specially designed and implemented to simulate the probe-based topology discovery mechanism in the presence of middleboxes, for testing the large scale networks. The inputs (list of middleboxes, their interfaces and route details) required to this tool are provided as a network configuration file.

\section{Probe Security \& protocol}
\begin{figure}[t]
\centering
\includegraphics{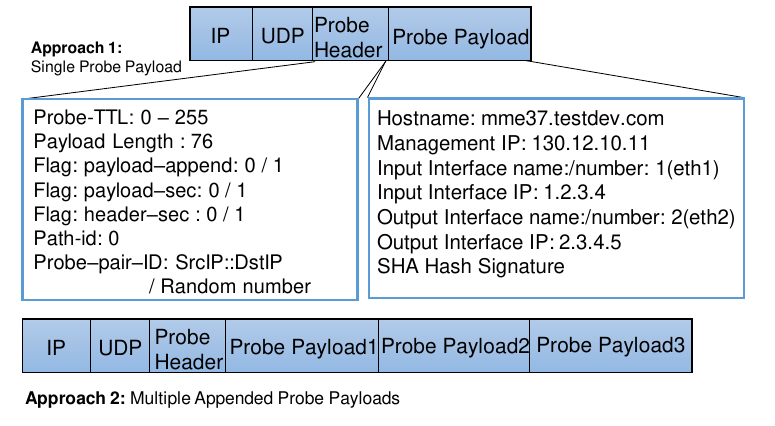}
\caption{Probe message header and payload details.}
\label{figure: Probe protocol details}
\end{figure}

\noindent \newline
\noindent In this section, we demonstrate security support to our probe mechanisms mentioned in Approaches 1 and 2 (figure \ref{figure: probe traversal}). By securing the probe header and payload, we protect it against the \textit{topology poisoning}, \textit{data leak} and \textit{data tampering} attacks. We limit our threat model to network eavesdropping, while the compromised middleboxes could still poison our network topology. The protocol details are shown in figure \ref{figure: Probe protocol details}.

\subsection{Probe Header details} 


\noindent The probe message header fields are detailed below,

\noindent \textbf{Probe-TTL.} At each hop the middlebox agent increments this field (counter) by 1, before forwarding it further and it is discarded when the Probe-TTL reaches the maximum threshold value configured by the \textit{Topology Manager}. This field helps in detecting both the sequence and the number of devices the probe traversed in the network. 

\noindent \textbf{Path-id.} The path-id field serves two different purposes, it identifies the type of probe messages i.e., \textit{discovery probe} (path-id=0) or \textit{path checker probe} (path-id>0) and also defines the path for the probe traversal. 

\noindent \textbf{PayloadLength.} This field is used for storing the probe payload size, which is updated at each hop.

\noindent \textbf{Flag:Payload-append.} This flag forces the MB Agent to append device details to the probe payload before forwarding it further, avoiding the up-call at each hop. 

\noindent \textbf{Probe-pair-ID.}
This header field carries either probe-pair IP addresses or the Random Number depending on the header-sec flag. 

\noindent \textbf{Flag: header-sec.} 
This field defines if the '\textit{probe-pair-ID}' header field carries the probe-pair IP address directly or else uses the Random Number  to hide the probe-pair details. Middleboxes along the path resolves this Random Number to the IP address of the probe-pair using the {\small 'RESOLVE-PROBEID'} API. 

\noindent \textbf{Flag: Payload-Sec.} This flag defines the encryption status of the payload.

\subsection{Payload security}
Our approach uses both the symmetric key and public key encryption mechanisms for securing the payload data. The encryption procedure is detailed below,

\begin{enumerate}
\item 
Initially, the probe payload is encrypted using the random AES symmetric key generated by the MB Agent inside the middlebox. 

\item
For providing data-integrity hash the encrypted data along with some of the important header fields (\textit{probe-pair-ID, Flag:header-sec, Flag:payload-sec and Flag:ProbeAppend}) using the SHA-256.

\item
The generated AES key and the hash calculated on the 'header and encrypted data' is again encrypted using the public key distributed by the MB Controller.

\item
The encrypted key and hash is concatenated with encrypted payload message and forwarded further.

\end{enumerate}

while the field \textit{payload-sec} defines the security of payload. Depending on the \textit{payload-append} field, the encrypted payload is either shared with the controller at each hop or appended till the destination probe-pair, when the MB controller receives the encrypted payload, the AES key and hash is extracted first using private key of the MB Controller. Later the AES key is used to decrypt the actual payload content. The MB Controller recalculates the hash on the probe and compares it with the hash values stored in the payload for data integrity. The payload security is tested with Nacl (salt)  \cite{NACL} integrated as part of both the MB Agent and MB Controller, this module is considered lighter with better latency parameters \cite{FASTERNACL} when compared with openssl. 

\begin{figure}[t]
\centering
\includegraphics[height=38mm]{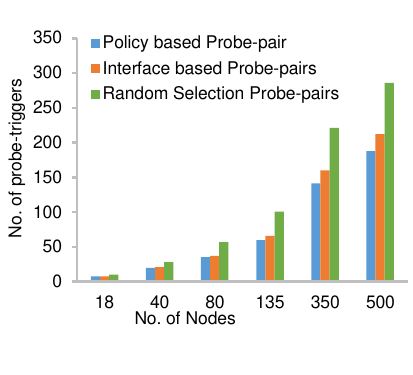}
\includegraphics[height=37mm]{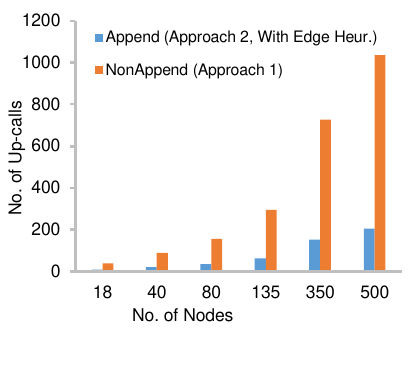}
\vskip -15pt
\caption{Topology discovery(probe-triggers, up-calls).}
\label{figure: graph12}
\end{figure} 

\begin{table}[t]
\begin{tabular}{ |l|c|c|c|c|c| }\hline
\multicolumn{1}{|c|}{\multirow{1}{*}{Topology}}	
&	\multicolumn{2}{c|}{Avg. Probe-triggers}	& \multicolumn{2}{c|}{Avg. Up-Calls} \\

\cline{2-5}

\ (20 Nodes)&Edge&Random&Append& Non-\\
&Heur.&Select&&Append\\
 \hline
Cisco Ent.  & 9.71 & 11.22 & 9.97 & 41.66\\
In-\&Offline   & 8 & 9.35 & 8.6 & 43.7\\
Tree           & 6.25 & 10.25 & 10.75 & 35.1\\
Full Mesh     & 187.5 & 195 & 192.3 & 403\\

\hline
\end{tabular}
\caption{Topologies vs probe-based discovery.}
\label{table1}
\end{table}

\section{IMPLEMENTATION}
\noindent \newline
\noindent The MB Agent module integrated into each of middlebox runs a UDP based transparent proxy integrated with NaCl (salt) \cite{NACL} and uses iptables (Netfilter) in PREROUTING of \textit{Probe-Handling} module. The \textit{Probe Tracking \& Control} module helps MB Agent in monitoring and steering the outgoing probe messages using iptables and ip rules. The MB Controller is also integrated with NaCl to handle the encrypted probe messages. 

\noindent \textbf{Compatibility.} The code changes that we introduced in middleboxes pertaining to probe routing doesn't impact any of the middlebox functionalities. As most of the third-party middlebox vendors such as Cisco~\cite{Cisco}, Juniper~\cite{Juniper} and Barracuda Networks~\cite{Barracuda} supports the port (transport) based routing functionality in their middleboxes, it will become easier for us to integrate our solution into the third-party middleboxes.

\noindent \textbf{Assumption.} The middleboxes are expected to be have simple infrastructure routing rules configured to handle probe messages.

\section{EVALUATIONS}
\noindent \newline
\noindent The results shown in figure \ref{figure: graph12} are evaluated considering the Cisco's medium scale enterprise network design \cite{Mediumscaledesignprofile} topologies with up-to 500 middleboxes and 60 switches, tested using the simulator tool and the figure shows the average number of probe-triggers (i.e., probe messages generated by the middleboxes to discover the topology) and up-calls generated for discovering the topology, considering both the heuristics described in section \ref{probe pair selection}. The table\ref{table1} shows the simulation results of different types of 20-Node topologies, with Mesh and In- and Offline topologies considered as boundary cases for our probe-based topology discovery mechanism. 

In lab environment, experiments are conducted with around 40 nodes integrated with our MB Agent on systems including third-party middleboxes such as RouteFinder \cite{ROUTEFINDER} and PC/Server systems. The PC based systems are also integrated with open-source middlebox function such as OpenSwan, bro, Squid, Netfilter, openvswitch and vconfig used as IPSec VPN, IDS, Web Proxy, Firewall, switching and VLAN functions respectively. We evaluated all our results with different topologies that are commonly used in any enterprise networks, with an average of around 2.1 to 2.7 interfaces per middlebox, with exception to the fully connected Mesh topology.

\noindent \textbf{Topology Discovery Time.} The average time involved in discovering Cisco's small scale enterprise network topology with 40 nodes is well with in 5 seconds and with Nacl (salt) \cite{NACL} integrated for probe security it has costed us around 8 sec for the same network.

\section{CONCLUSION \& FUTURE WORK} \label{conclusion}

\noindent \newline
\noindent In this paper, we presented an Unified infrastructure framework and obtained global visibility of enterprise network in the presence of middleboxes, with key focus on topology discovery and end-to-end path connectivity verification. Though this topology discovery mechanism is tested specifically to enterprise network device deployment scenarios, we believe it could be easily extended to other scenarios and types of networks.

As part of our future work, we (i) plan to integrate our implementation with Openstack and test it on large scale test environments. (ii) analyze the overhead involved in different security mechanisms used for topology discovery. 
(iii) reduce discovery time with PROBE-INIT parallelization.



\end{document}